\documentclass[11pt]{article}
\usepackage{moriond,epsfig}

\newcommand{\rt}{\rightarrow}

\def\Journal#1#2#3#4{{#1} {\bf #2}, #3 (#4)}

\def\NIM{\em Nucl. Instrum. Methods}

\def\NPB{{\em Nucl. Phys.} B}
\def\PLB{{\em Phys. Lett.}  B}
\def\PRL{\em Phys. Rev. Lett.}
\def\PRD{{\em Phys. Rev.} D}

\def\NPA{{\em Nucl. Phys.} A}

\def\be{\begin{equation}}
\def\ee{\end{equation}}
\def\bea{\begin{eqnarray}}
\def\eea{\end{eqnarray}}

\begin{document}
\vspace*{4cm}
\title{NEW $R$ VALUES IN 2-5 GeV FROM THE BEIJING SPECTROMETER}
\author{ G. S. HUANG representing the BES Collaboration }
\address{Institute of High Energy Physics, 19 Yuquan Road, Beijing 100039,
China\\E-mail: huanggs@pony1.ihep.ac.cn}

\maketitle\abstracts{
The values of $R = \sigma(e^+e^-\rightarrow
\mbox{hadrons})/\sigma(e^+e^-\rightarrow\mu^+\mu^-)$ for 85
center-of-mass energies between 2 and 5 GeV were measured 
with the upgraded Beijing Spectrometer at the Beijing 
Electron-Positron Collider, with an average uncertainty of $\sim7\%$.}

\section{Introduction}
The QED running coupling constant evaluated at the $Z$ pole, 
$\alpha(M^2_{Z})$, and the anomalous 
magnetic moment of the muon, $a_{\mu}=(g-2)/2$, are two fundamental 
quantities that are used to test the Standard Model (SM)~\cite{rlowe}. 
The dominant uncertainties in both $\alpha(M^2_{Z})$ and $a_{\mu}^{SM}$ are 
due to the effects of hadronic vacuum polarization, which cannot be 
reliably calculated.  Instead, with the application of dispersion relations,
experimentally measured $R$ values are used to determine the vacuum 
polarization~\cite{rlowe}, where $R$ is the lowest order cross section 
for $e^+e^-\rightarrow\gamma^*\rightarrow \mbox{hadrons}$
in units of the lowest-order QED cross section for
$e^+e^- \rightarrow \mu^+\mu^-$, namely
$R=\sigma(e^+e^- \rightarrow \mbox{hadrons})/\sigma(e^+e^-\rightarrow
\mu^+\mu^-)$, where 
$\sigma (e^+e^- \rightarrow \mu^+\mu^-) = \sigma^0_{\mu \mu}=
4\pi \alpha^2(0) / 3s$.
   
The uncertainties in $\alpha(M^2_{Z})$ and
$a^{SM}_{\mu}$ are dominated by the errors in the 
values of $R$ in the center of mass (cm) energy range below 5 GeV.  These were
measured about 20 years ago with a precision of about $15\sim 20\%$.
Thus, new measurements of $R$ in the energy region between 2 and 5 GeV
with significantly improved precision are very important~\cite{rlowe}.  
In this paper, we report measurements of $R$ at 85
cm energies between 2 and 5 GeV, with an average precision
of $6.6\%$.

\section{$R$ scan with BESII at BEPC}
The measurements were carried out using the upgraded Beijing
Spectrometer (BESII)~\cite{bes} at the Beijing Electron 
Positron Collider (BEPC). 
BESII is a conventional collider detector based on
a large solenoid magnet with a central field of 0.4~T.

Following a preliminary scan that measured $R$ at six energy points
between 2.6 and 5 GeV~\cite{besr_1}, we performed a finer 
$R$ scan with 85 energy points covering
the energy region between 2 and 4.8 GeV~\cite{rscan2}. 
In order to understand beam-associated backgrounds,
separated beam data were accumulated at 24 different energies and single
beam runs for both $e^-$ and $e^+$ were done at 7 energies
interspersed throughout the entire energy range. Special runs
were taken at the $J/\psi$ resonance to determine the trigger efficiency. 
The $J/\psi$ and $\psi(2S)$ resonances were scanned at the beginning and
end of the $R$ scan to calibrate the cm energy.

\section{Data Analysis} 
Experimentally, the value of $R$ is determined from the number of 
observed hadronic events, $N^{obs}_{had}$, by the relation
\begin{equation}
R=\frac{ N^{obs}_{had} - N_{bg} - \sum_{l}N_{ll} - N_{\gamma\gamma} }
{ \sigma^0_{\mu\mu} \cdot L \cdot \epsilon_{had} \cdot \epsilon_{trg}
\cdot (1+\delta)},
\end{equation}
where $N_{bg}$ is the number of beam-associated background events;
$\sum_{l}N_{ll},~(l=e,\mu,\tau)$ are the numbers
of lepton-pair events from one-photon processes and $N_{\gamma\gamma}$
the number of two-photon process
events that are misidentified as hadronic events;
$L$ is the integrated luminosity; $\delta$ is
the radiative correction; $\epsilon_{had}$ is the detection efficiency 
for hadronic events; and $\epsilon_{trg}$ is the trigger efficiency.

The trigger efficiencies, measured by comparing the
responses to different trigger requirements
in special runs taken at the $J/\psi$ resonance,
are determined to be 99.96\%, 99.33\% and 99.76\%
for Bhabha, dimuon and hadronic events, respectively, with
an error of 0.5\%~\cite{besr_1}.

We developed a set of 
requirements on fiducial regions, vertex positions, track fit 
quality, maximum and minimum BSC energy deposition, track momenta and 
time-of-flight hits that preferentially distinguish one-photon
multi-hadron production from all possible contamination
mechanisms.
Residual background contributions are due
to cosmic rays, lepton pair production, two-photon
interactions and single-beam-related processes. Additional
requirements are imposed on two-prong events, for which
cosmic ray and lepton pair backgrounds are especially
severe~\cite{besr_1}.

The number of hadronic events and the beam-associated background level 
are determined by fitting the distribution of event vertices along the 
beam direction with a Gaussian for real hadronic events and a 
polynomial of degree two for the background, which
can also be subtracted by applying the 
same hadronic event selection criteria to separated-beam data~\cite{besr_1}.
The differences in $R$ values with these two methods
range between 0.3 and 2.3\%,
depending on the energy.

The integrated luminosity is determined
from the number of large-angle Bhabha events selected
using only the BSC energy deposition. 

JETSET, the commonly used event generator 
for $e^+e^-\rt \mbox{hadrons}$, was not 
intended to be applicable to the low
energy region, especially that below 3 GeV. 
A new generator, LUARLW, was developed by the Lund 
group and the BES collaboration.
The generator uses a formalism based on the 
Lund Model Area Law, but without the extreme-high-energy 
approximations used in JETSET's string fragmentation algorithm~\cite{bo}. 
The final states simulated in LUARLW are exclusive 
in contrast to JETSET, where they are inclusive. 
Above 3.77 GeV, the production of charmed mesons 
is included in the generator according to the Eichten 
Model~\cite{eichiten,chenjc}.

The parameters in LUARLW are tuned to reproduce
the observed multiplicity, sphericity, angular and momentum 
distributions, etc.,  over the entire energy region covered by the scan.
The uncertainty of detection efficiency is 2\%, estimated by varying the
parameters in LUARLW.
The detection efficiencies were also determined using JETSET74 for 
the energies above 3 GeV. 
The difference between the JETSET74 and LUARLW results is about 1\%,
and is also taken into account in estimating the systematic uncertainty.    

Different schemes for the radiative corrections were compared 
\cite{berends,fmartin,fadin,crystalball}.
The schemes of Refs.~\cite{fadin} and \cite{crystalball} take into 
account vacuum polarization not only for electrons and muons, but also 
taus and hadrons.  The correction factors calculated with these two
approaches are consistent within 0.5\% in the continuum and 
differ by less than 1\% in the charm resonance region. 
The formalism of Ref. \cite{crystalball} is used in our calculation, and
differences with the schemes described in Ref. \cite{fadin} are 
included in the systematic errors. In the calculation of the
radiative correction above charm threshold,
where the resonances are broad and where the total width of the resonance
is energy-dependent, we take the interference between resonances into
account.

\section{The Results}
\begin{figure}[htb] 
\epsfysize=3.9in
\centerline{\epsfbox{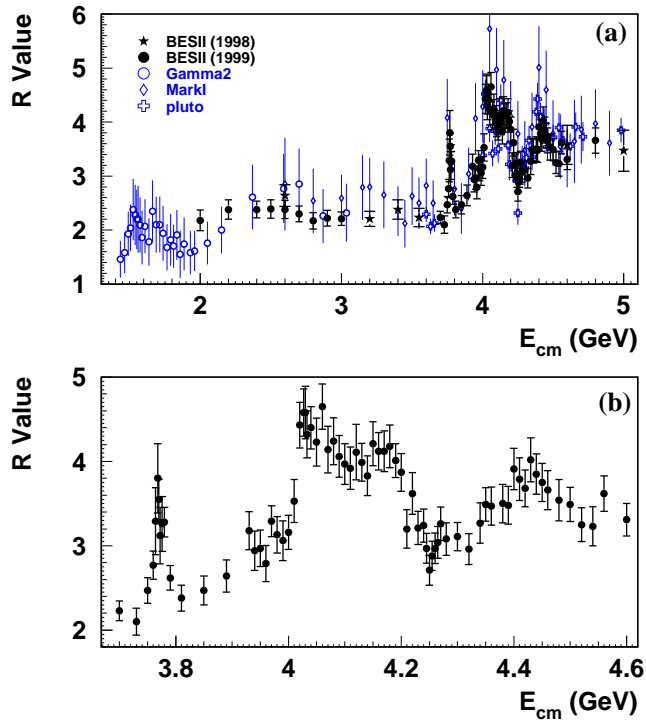}}
\caption{(a) A compilation of measurements of $R$ in the cm
energy range from 1.4 to 5 GeV. (b) $R$ values from this experiment 
in the resonance region between 3.75 and 4.6 GeV.} 
\label{fig:besr}
\end{figure}

Table~\ref{tab:rvalue} lists the $R$ values measured 
by BES in this experiment. They are displayed in 
Fig.~\ref{fig:besr}, together with BESII values from Ref.~\cite{besr_1} and 
those measured by MarkI,
$\gamma\gamma 2$, and Pluto~\cite{mark1,gamma2,pluto}.
The $R$ values from BESII have an average uncertainty of
about 6.6\%, which represents
a factor of two to three improvement in precision 
in the 2 to 5 GeV 
energy region.  These improved measurements should have a significant impact on the 
global fit to the electroweak data and
the determination of the SM prediction for the mass
of the Higgs particle.   In addition, they
are expected to provide an
improvement in the precision of the calculated value of
$a_{\mu}^{SM}$~\cite{ichep2k,martin}.

\begin{table*}[htbp]
\begin{center}
\caption{The measured $R$ values obtained in this
experiment; the first error is statistical, the second systematic.} 
\begin{scriptsize}
\begin{tabular}{cccccccc} \hline
$E_{cm}$ & $R$   & $E_{cm}$ & $R$ &   $E_{cm}$ & $R$ &
$E_{cm}$ & $R$ \\
(GeV)& & (GeV)& & (GeV)& & (GeV)& \\ \hline
2.000& $2.18\pm0.07\pm0.18$ &3.890& $2.64\pm0.11\pm0.15$ &4.120
& $4.11\pm0.24\pm0.23$ &4.340& $3.27\pm0.15\pm0.18$ \\
2.200& $2.38\pm0.07\pm0.17$ &3.930& $3.18\pm0.14\pm0.17$ &4.130
& $3.99\pm0.15\pm0.17$ &4.350& $3.49\pm0.14\pm0.14$ \\
2.400& $2.38\pm0.07\pm0.14$ &3.940& $2.94\pm0.13\pm0.19$ &4.140
& $3.83\pm0.15\pm0.18$ &4.360& $3.47\pm0.13\pm0.18$ \\
2.500& $2.39\pm0.08\pm0.15$ &3.950& $2.97\pm0.13\pm0.17$ &4.150
& $4.21\pm0.18\pm0.19$ &4.380& $3.50\pm0.15\pm0.17$ \\
2.600& $2.38\pm0.06\pm0.15$ &3.960& $2.79\pm0.12\pm0.17$ &4.160
& $4.12\pm0.15\pm0.16$ &4.390& $3.48\pm0.16\pm0.16$ \\
2.700& $2.30\pm0.07\pm0.13$ &3.970& $3.29\pm0.13\pm0.13$ &4.170
& $4.12\pm0.15\pm0.19$ &4.400& $3.91\pm0.16\pm0.19$ \\
2.800& $2.17\pm0.06\pm0.14$ &3.980& $3.13\pm0.14\pm0.16$ &4.180
& $4.18\pm0.17\pm0.18$ &4.410& $3.79\pm0.15\pm0.20$ \\
2.900& $2.22\pm0.07\pm0.13$ &3.990& $3.06\pm0.15\pm0.18$ &4.190
& $4.01\pm0.14\pm0.14$ &4.420& $3.68\pm0.14\pm0.17$ \\
3.000& $2.21\pm0.05\pm0.11$ &4.000& $3.16\pm0.14\pm0.15$ &4.200
& $3.87\pm0.16\pm0.16$ &4.430& $4.02\pm0.16\pm0.20$ \\
3.700& $2.23\pm0.08\pm0.08$ &4.010& $3.53\pm0.16\pm0.20$ &4.210
& $3.20\pm0.16\pm0.17$ &4.440& $3.85\pm0.17\pm0.17$ \\
3.730& $2.10\pm0.08\pm0.14$ &4.020& $4.43\pm0.16\pm0.21$ &4.220
& $3.62\pm0.15\pm0.20$ &4.450& $3.75\pm0.15\pm0.17$ \\
3.750& $2.47\pm0.09\pm0.12$ &4.027& $4.58\pm0.18\pm0.21$ &4.230
& $3.21\pm0.13\pm0.15$ &4.460& $3.66\pm0.17\pm0.16$ \\
3.760& $2.77\pm0.11\pm0.13$ &4.030& $4.58\pm0.20\pm0.23$ &4.240
& $3.24\pm0.12\pm0.15$ &4.480& $3.54\pm0.17\pm0.18$ \\
3.764& $3.29\pm0.27\pm0.29$ &4.033& $4.32\pm0.17\pm0.22$ &4.245
& $2.97\pm0.11\pm0.14$ &4.500& $3.49\pm0.14\pm0.15$ \\
3.768& $3.80\pm0.33\pm0.25$ &4.040& $4.40\pm0.17\pm0.19$ &4.250
& $2.71\pm0.12\pm0.13$ &4.520& $3.25\pm0.13\pm0.15$ \\
3.770& $3.55\pm0.14\pm0.19$ &4.050& $4.23\pm0.17\pm0.22$ &4.255
& $2.88\pm0.11\pm0.14$ &4.540& $3.23\pm0.14\pm0.18$ \\
3.772& $3.12\pm0.24\pm0.23$ &4.060& $4.65\pm0.19\pm0.19$ &4.260
& $2.97\pm0.11\pm0.14$ &4.560& $3.62\pm0.13\pm0.16$ \\
3.776& $3.26\pm0.26\pm0.19$ &4.070& $4.14\pm0.20\pm0.19$ &4.265
& $3.04\pm0.13\pm0.14$ &4.600& $3.31\pm0.11\pm0.16$ \\
3.780& $3.28\pm0.12\pm0.12$ &4.080& $4.24\pm0.21\pm0.18$ &4.270
& $3.26\pm0.12\pm0.16$ &4.800& $3.66\pm0.14\pm0.19$ \\
3.790& $2.62\pm0.11\pm0.10$ &4.090& $4.06\pm0.17\pm0.18$ &4.280
& $3.08\pm0.12\pm0.15$ &     &                      \\
3.810& $2.38\pm0.10\pm0.12$ &4.100& $3.97\pm0.16\pm0.18$ &4.300
& $3.11\pm0.12\pm0.12$ &     &                      \\
3.850& $2.47\pm0.11\pm0.13$ &4.110& $3.92\pm0.16\pm0.19$ &4.320
& $2.96\pm0.12\pm0.14$ &     &                      \\ \hline
\end{tabular}
\end{scriptsize}
\label{tab:rvalue}
\end{center}
\end{table*}

Further improvements in the accuracy of $R$ measurements 
at BEPC will require higher machine luminosity, especially 
for energies below 3.0 GeV, and better detector performance, 
particularly in the area of calorimetry. 
Increased precision in the areas of 
hadronic event simulation and the calculation
of the radiative correction are also  required.

\section*{Acknowledgments}
We would like to thank the staff of the BEPC Accelerator Center
and IHEP Computing Center for their efforts.  
We thank B. Andersson for helping in the development of the LUARLW generator.
We also wish to acknowledge useful discussions with M. Davier, 
B. Pietrzyk, T. Sj\"{o}strand,  A. D. Martin and M. L. Swartz.
We especially thank M. Tigner for major contributions not only to
BES but also to the operation of the BEPC during the $R$ scan.

This work is supported in part by the National Natural
 Science Foundation of China under Contract Nos. 19991480,
19805009 and 19825116; the Chinese
 Academy of Sciences under contract Nos. KJ95T-03, and E-01 (IHEP);
 and by the Department of
 Energy under Contract Nos.
 DE-FG03-93ER40788 (Colorado State University),
 DE-AC03-76SF00515 (SLAC),
 DE-FG03-94ER40833 (U Hawaii), DE-FG03-95ER40925 (UT Dallas),
and by the Ministry of Science and Technology of Korea under Contract
KISTEP I-03-037(Korea).

\end{document}